\documentclass[prl,twocolumn,showpacs,amsmath,amssymb,superscriptaddress]{revtex4}
\usepackage{graphicx}
\usepackage{multirow}
\usepackage{dcolumn}
\usepackage{amssymb,amscd,xypic,bm,dsfont,wasysym}
\usepackage{float}
\usepackage{color}
\usepackage{xcolor}

\newcommand{\footstar}[1]{$^{**}$\footnotetext{$^{**}$#1}}
\begin{document}

\title{Topological nature of FeSe$_{0.5}$Te$_{0.5}$ superconductor }

\author{Zhijun Wang$^{**}$,$^{1,3}$
  P. Zhang\footstar{these authors contributed equally to this work.},$^1$
Gang Xu,$^{1,4}$ L.K. Zeng,$^1$ H. Miao,$^1$ Xiaoyan Xu,$^1$ T. Qian,$^1$ \\
Hongming Weng,$^{1,2}$ P. Richard,$^1$ A. V. Fedorov,$^5$
H. Ding}
\email{dingh@aphy.iphy.ac.cn}

\author{Xi Dai}
\email{daix@aphy.iphy.ac.cn}

\author{Zhong Fang}
\email{zfang@aphy.iphy.ac.cn}

\affiliation{Beijing National Laboratory for Condensed Matter
  Physics, and Institute of Physics, Chinese Academy of Sciences,
  Beijing 100190, China}
\affiliation{Collaborative Innovation Center of Quantum Matter, Beijing, China \\
 $^3$ Department of Physics, Princeton University, Princeton NJ 08544, USA\\
 $^4$Department of Physics, McCullough Building,
     Stanford University, Stanford CA 94305-4045, USA\\
 $^5$Advanced Light Source, Lawrence Berkeley
      National Laboratory, Berkeley, California 94720, USA}

\date{\today}

\begin{abstract}
  We demonstrate, using first-principles calculations, that the
  electronic structure of FeSe$_{1-x}$Te$_{x}$ ($x$=0.5) is
  topologically non-trivial, characterized by an odd $\mathbb Z_2$
  invariant and Dirac cone type surface states, in sharp contrast to
  the end member FeSe ($x$=0). This topological state is induced
  by the enhanced three-dimensionality and spin-orbit coupling due
  to Te substitution (compared to FeSe), characterized by a band
  inversion at the $Z$ point of the Brillouin zone, which is confirmed
  by our ARPES measurements. The results suggest that the surface of
  FeSe$_{0.5}$Te$_{0.5}$ may support a non-trivial
  superconducting channel in proximity to the bulk.  
\end{abstract}

\pacs{73.43.-f, 74.70.Xa, 73.20.-r}

\maketitle

\begin{figure}[tb]
\includegraphics[clip,scale=0.48]{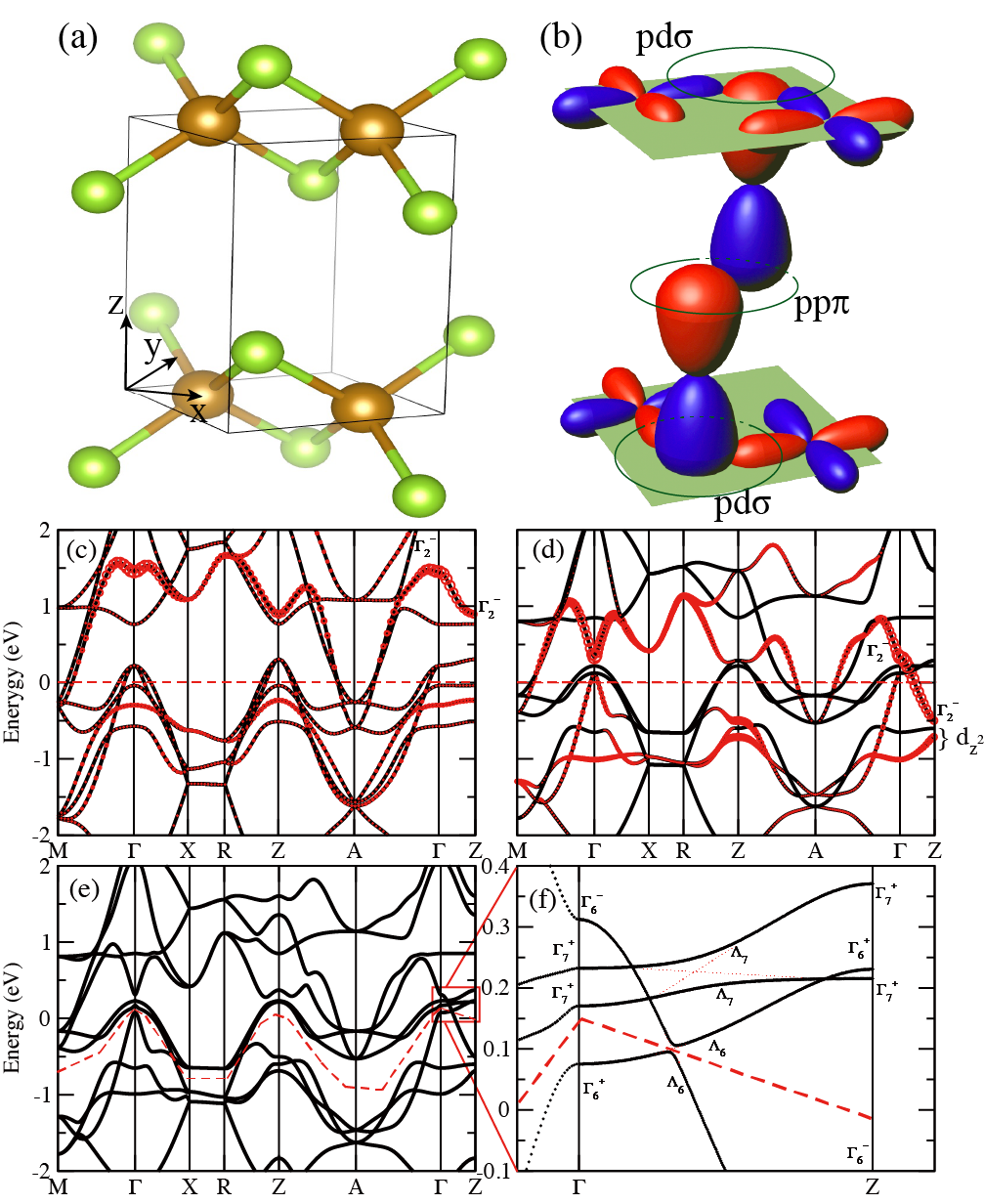}
\caption{(Color online)
 The crystal structure and DFT electronic band structures of FeX (X = Se$_{0.5}$Te$_{0.5}$).
 (a) is the crystal structure with the $x$-axis pointing along Fe nearest neighbors. Grey and green balls represent Fe and X atoms, respectively. (b) Schematic plot of the hybridization
  along the $z$ axis between the combined orbitals $D_{xy}^-$ and $P_z^-$ (see details in appendix B), consisting of intralayer $pd\sigma$ and interlayer $pp\pi$ hopping.
(c) Band structure of FeSe with internal parameter z$_X$=0.2345 without SOC. (d) Band structure of FeSe$_{0.5}$Te$_{0.5}$ (z$_X$=0.2719) without SOC. (e) Band structure of FeSe$_{0.5}$Te$_{0.5}$ with SOC. (f) Zoom-in view of the solid red box area in (e). The size of red the circles in (c) and (d) indicates the weight of the $p_z$ component of the chalcogen atoms. The two $d_{z^2}$ bands are indicated below $E_F$ in (d). The original $\Lambda_5$ bands split into $\Lambda_6$ and $\Lambda_7$. Along the $\Gamma$-Z line, two $\Lambda_6$ bands cross and hybridize to open a SOC gap of about 10 meV. The red dashed line corresponds to a Fermi curve across the gap.
}\label{str}
\end{figure}

Among the Fe-based superconductors, the FeSe$_{1-x}$Te$_x$ family of 
compounds~\cite{FeSe,doping1,doping2,doping3} is of particular interest. First, it has the simplest PbO
structure (space group $P4/nmm$) with Se (or Te) atoms forming
distorted tetrahedra around Fe (see Fig.~\ref{str}(a)) similar to the structure
of FeAs planes in the families of FeAs-based high $T_c$
superconductors~\cite{LaOFeAs}. Second, the internal parameters can be systematically
tuned by the substitution of Se by Te~\cite{lisl,xxwu,fst2015}, which provides us a platform for in-depth study of possible superconducting mechanisms and topological characters.
Thirdly,
superconductivity has been observed for a wide range of
composition $x$~\cite{doping1,doping2,doping3}, and the transition 
temperature $T_c$ can be further enhanced by pressure~\cite{presure1,presure2,presure3}.
 More recently, superconductivity with $T_c$
higher than 77 K was suggested for single unit cell FeSe films~\cite{fese77}
epitaxially grown on SrTiO$_3$ substrates.

Despite these interesting properties though, the particularities of the
system have still not been fully explored. Earlier studies, both
theoretical and experimental, suggest the similarity of the
electronic structures of the Fe chalcogenides (FeSe, FeTe)~\cite{ldasete,dmft,tamai2010} and the
FeAs-based~\cite{ldafeas,xu2008,haule2008} superconductors. Indeed, the low-energy physics around the
Fermi level is dominated by the Fe-$3d$ states, and the morphology of
the Fermi surfaces is similar. On the other hand, a surprisingly stable (no splitting under external magnetic field) zero-energy bound state
(ZBS) at randomly distributed interstitial excess Fe sites was observed in very recent
scanning tunneling microscopy (STM) measurements on the surface of superconducting Fe(Te,Se)~\cite{zeropeak}, suggesting possible topological feature of its electronic structure. Obviously, the $5p$ orbitals of Te are more extended and
have stronger spin-orbit coupling (SOC) than the $4p$ orbitals of Se. The consequences of Te
substitution, particularly for the bulk topological character of
FeSe$_{1-x}$Te$_x$, have been largely ignored in the literature and will
be the main purpose of the present paper. Based on 
first-principles calculations combined with angle resolved photoemission spectroscopy (ARPES) measurements, here we report that the electronic structure of FeSe$_{0.5}$Te$_{0.5}$ is
topologically non-trivial, in sharp contrast to its end member FeSe.
The topological properties of FeSe$_{0.5}$Te$_{0.5}$ can be
characterized by an odd $\mathbb Z_2$ number, and the existence of Dirac cone
type surface states, in proximity to bulk superconductivity,
should favor topologically superconducting surface states, as suggested
by Fu and Kane~\cite{fu2008}.


The electronic structures of FeSe and FeSe$_{0.5}$Te$_{0.5}$ are
calculated with SOC included. The calculations are performed based on
the density functional theory (DFT) ~\cite{HK1964,KS1965} and the generalized gradient
approximation (GGA) for the exchange-correlation potential~\cite{ggapbe}, as
implemented in the plane-wave pseudopotential based BSTATE (Beijing Simulation Tool of Atomic TEchnology) package~\cite{fangcode}. 
The experimental lattice parameters~\cite{lisl} are used
in the calculations. Maximally localized Wannier functions (MLWF)~\cite{SMV} are
constructed from bulk calculations, and then used to study the surface
states in the semi-infinite system. To treat the substitution
properly, we have calculated FeSe$_{0.5}$Te$_{0.5}$ by using both
the virtual crystal approximation and the two-formula cell with ordered Se
and Te sites. Both calculations give converging results.

We first neglect the SOC and concentrate on the comparison between the
electronic structures of FeSe and FeSe$_{0.5}$Te$_{0.5}$ (as
shown in Fig.~\ref{str} (c) and (d)). The band structure of FeSe is very
similar to that of LaOFeAs as reported before~\cite{xu2008}. At the $\Gamma$ point,
the valence band top is not occupied, leading to the well known hole
pockets of Fermi surfaces around $\Gamma$. The three top-most states
at $\Gamma$ can be labeled as two-fold degenerate $\Gamma_5^+$
states ($d_{yz}$/$d_{xz}$ orbitals) and non degenerate $\Gamma_4^+$ state
($d_{xy}$ orbital), respectively. There is a clear band gap larger than 0.6 eV above the valence band top at the $\Gamma$
point. All the $d-d$ anti-bonding states (with negative parity) are located above the
gap except the $d_{z^2}$ orbital that has a weaker anti-bonding state.
Among them, the most interesting state is the second highest one
with remarkable red circles, which belongs to the $\Gamma_2^-$
representation and comes from the anti-bonding $d_{xy}$ orbitals of
Fe and $p_z$ orbital of chalcogen. 
The weight of the $p_z$ component is illustrated by the size of red circles,
which suggests that the $\Gamma_2^-$ state can be affected by hybridizing with the $d_{xy}$ and $p_z$ orbitals. Looking along the $\Gamma-Z$ direction, the band dispersion of the
$\Gamma_2^-$ state is the strongest among all $d$ states.
Nevertheless, since it is energetically high, the band structure
around the $Z$ point of the Brillouin zone (BZ) is only slightly affected and remains similar
to that around $\Gamma$. The result suggests that FeSe is quite
two-dimensional with strong anisotropy.

Immediate differences can be seen in comparing these results with the band structure of FeSe$_{0.5}$Te$_{0.5}$ shown in
Fig.~\ref{str}(d): (1) the $\Gamma_2^-$ state in FeSe$_{0.5}$Te$_{0.5}$ is significantly pushed down and
almost touches the valence band top, and the band gap at $\Gamma$ (above
the valence and top) is nearly closed; (2) the band dispersion of this
$\Gamma_2^-$ state along the $\Gamma-Z$ direction is strongly
enhanced. As a result, a band inversion occurs at the $Z$ point, which implies a change of
topological property.


Compared to FeSe, the $c$-axis and the experimental intra-layer distance $d_z$ of FeSe$_{0.5}$Te$_{0.5}$ are enlarged by 8.2\% and 11\%,
respectively (See appendix A), while the $a$-axis and the inter-layer distance change only very little. As a result, the intra-layer
hybridizations, especially the $pd\sigma$ hopping as shown in Fig.~\ref{str}(b), are seriously weakened, and the $\Gamma_2^-$ band center is lowered very much and becomes close to $E_F$, as shown in Fig.~\ref{str}(d). On the other hand, because the Te $5p$
orbitals are much more extended than the Se $4p$ orbitals, the inter-layer hybridization through the $pp\pi$ bonding becomes stronger.
Therefore, the Te substitution enhances the hopping between layers and gives rise to a larger dispersion for $\Gamma_2^-$ in
FeSe$_{0.5}$Te$_{0.5}$. These two observations are in good agreement with the band structure of FeSe$_{0.5}$Te$_{0.5}$ shown in
Fig.~\ref{str}~(d). Therefore, the band inversion in FeSe$_{0.5}$Te$_{0.5}$ is a consequence of the weakened intra-layer hopping
and enhanced inter-layer hopping originating from the Te substitution.

Once SOC is included, the band structure of FeSe$_{0.5}$Te$_{0.5}$
opens a direct SOC gap and a non-trivial $\mathbb Z_2$ invariant can
be defined by assuming a ``curved chemical potential"-- the red dashed line
in Fig.~\ref{str}(e)--lying between the 10th and 11th bands (neglecting the spin-doublet degeneracy of the bands).
Generally, the doubly degenerate
$\Gamma_5^+$ states split into $\Gamma_6^+$ and $\Gamma_7^+$, and the
odd $\Gamma_2^-$ state turns into $\Gamma_6^-$ (see details in appendix B). 
Along the $\Gamma$-$Z$ high-symmetry line, 
two $\Lambda_6$ bands under C$_{4v}$ symmetry
hybridize and open a gap of about 10 meV, which is
clearly shown in Fig.~\ref{str}(f). When defining a Fermi curve
through the SOC gap, the $\mathbb{Z}_2$ invariant is easily calculated
from the parity criterion, which comes out to 1. (The parities at all
time-reversal invariant momenta (TRIM) are presented in
appendix B.) This non-zero $\mathbb{Z}_2$ invariant indicates
that FeSe$_{0.5}$Te$_{0.5}$ is in a topological phase that can support
the non-trivial surface states (SS). Due to the substantial SOC of Te, increasing
the content $x$ enlarges the SOC gap, which is beneficial for the
detection of the SS in the gap. Moreover, we also performed 
dynamical mean-field theory (DMFT)
calculations to confirm the band inversion and identify the strong
band renormalization due to electronic correlations (see
appendix C for details). The correlation effects do not
change the detail of the electronic bands, but simply reduce the
bandwidth. 

\begin{figure}[tb]
\includegraphics[clip,scale=0.40]{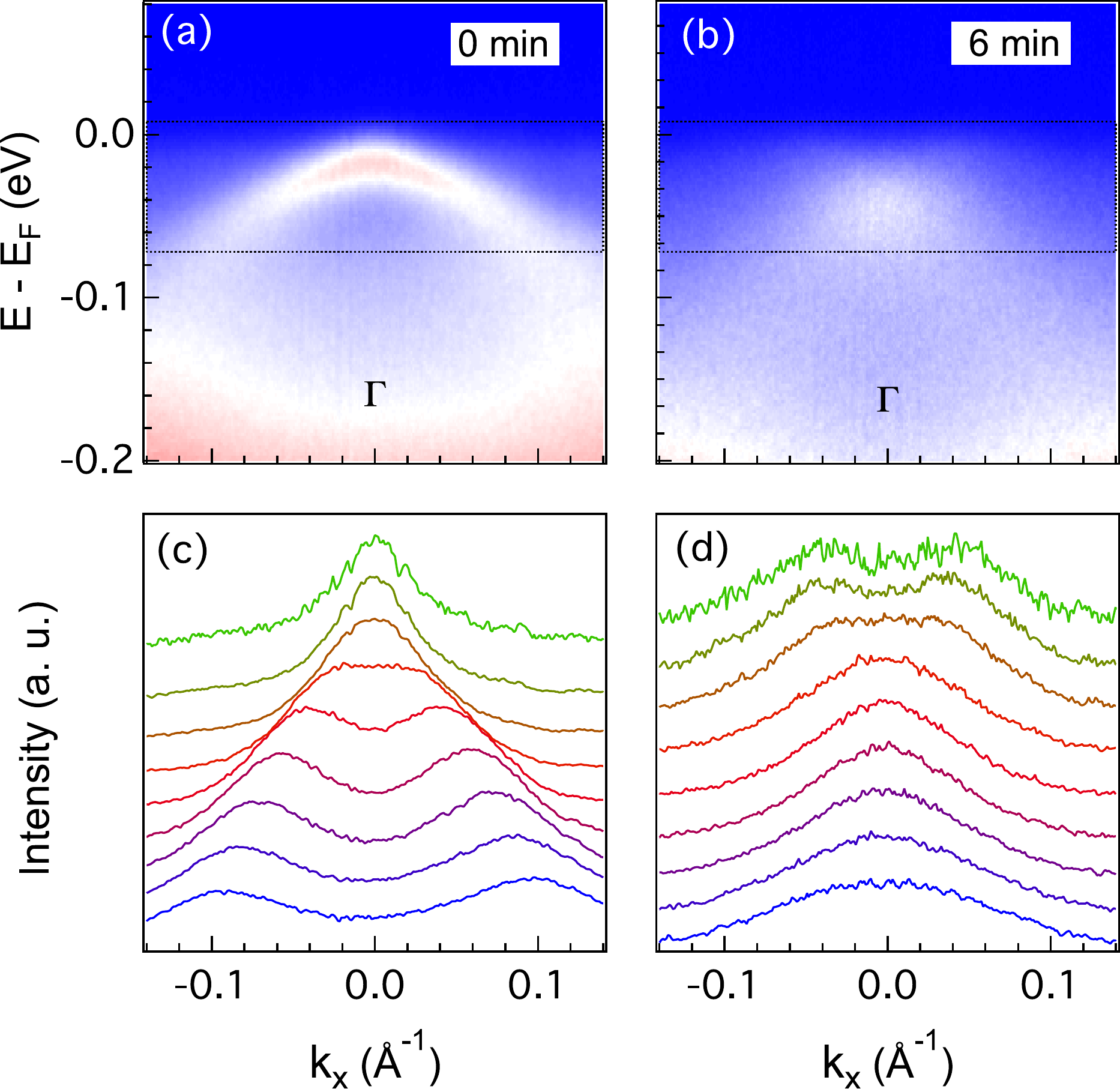}
 \caption{(Color online)
 ARPES spectra along the $\Gamma$-M direction in the $\sigma$ polarization geometry. (a) FeSe$_{0.5}$Te$_{0.5}$ before surface evaporation. (b) FeSe$_{0.5}$Te$_{0.5}$ after 6 minutes of K surface doping. (c), (d) MDCs corresponding to data from the dash boxes in panels (a) and (b), respectively.
 }\label{arp1}
\end{figure}


We performed ARPES measurements in order to demonstrate experimentally the existence of the $\Gamma_2^-$ band with strong $p_z$ orbital character
crossing $E_F$ along the $\Gamma$-Z direction. 
Large single-crystals of FeSe$_{0.45}$Te$_{0.55}$ were grown using the self-flux method, and LiFeAs with FeAs flux method. ARPES measurements were performed at the Advanced Light Source and Synchrotron Radiation Center, using a VG-Scienta electron analyzer. The light used was linearly polarized in directions parallel to the analyzer slit. The K source used for evaporation is made of a SAES K dispenser. In the experiments, the largest coverage is less than one mono-layer.
 All data were recorded with linear horizontal polarized photons  with a vertical analyzer slit ($\sigma$ geometry).
Under this configuration, odd orbitals with respect to the emission plane are visible, such as the $d_{xy}$ and $d_{yz}$ orbitals, while the $d_{xz}$ band should not be detected. In addition, our experimental setup leads to a $z$ component of the light polarization, and thus orbitals with mainly $z$ components such as $d_{z^2}$ and $p_z$ orbitals are also observable~\cite{liu2012}.

The comparison of the DFT band calculations on FeSe and FeSe$_{0.5}$Te$_{0.5}$, shown in Figs.~\ref{str}(c) and \ref{str}(d),
suggests that while it is pushed far above $E_F$ at $\Gamma$ in FeSe, the $\Gamma_2^-$ band in FeSe$_{0.5}$Te$_{0.5}$ forms an electron band just above $E_F$. To prove its existence, we raised the chemical potential by performing \emph{in-situ} K doping, and we measured ARPES spectra along $\Gamma$-M at h$\nu=30$ eV, which coincides with $k_z = 0$~\cite{zp2014}. The results before and after evaporation (6 minutes) are shown in Figs.~\ref{arp1}(a) and \ref{arp1}(b), respectively.  As expected, the hole band is sinking further below $E_F$ after evaporation. Interestingly, an additional electron band is observed, as clearly shown by contrasting the momentum distribution curves (MDCs) obtained before (Fig.~\ref{arp1}(c)) and after (Fig.~\ref{arp1}(d)) evaporation. This band, locating about 30 meV above the top of the valence band in FeSe$_{0.5}$Te$_{0.5}$, is very similar to the small 3D electron pocket reported in (Tl,Rb)$_y$Fe$_{2-x}$Se$_2$~\cite{liu2012}, which mainly has a $p_z$ component.

\begin{figure}[tb]
\includegraphics[clip,scale=0.40]{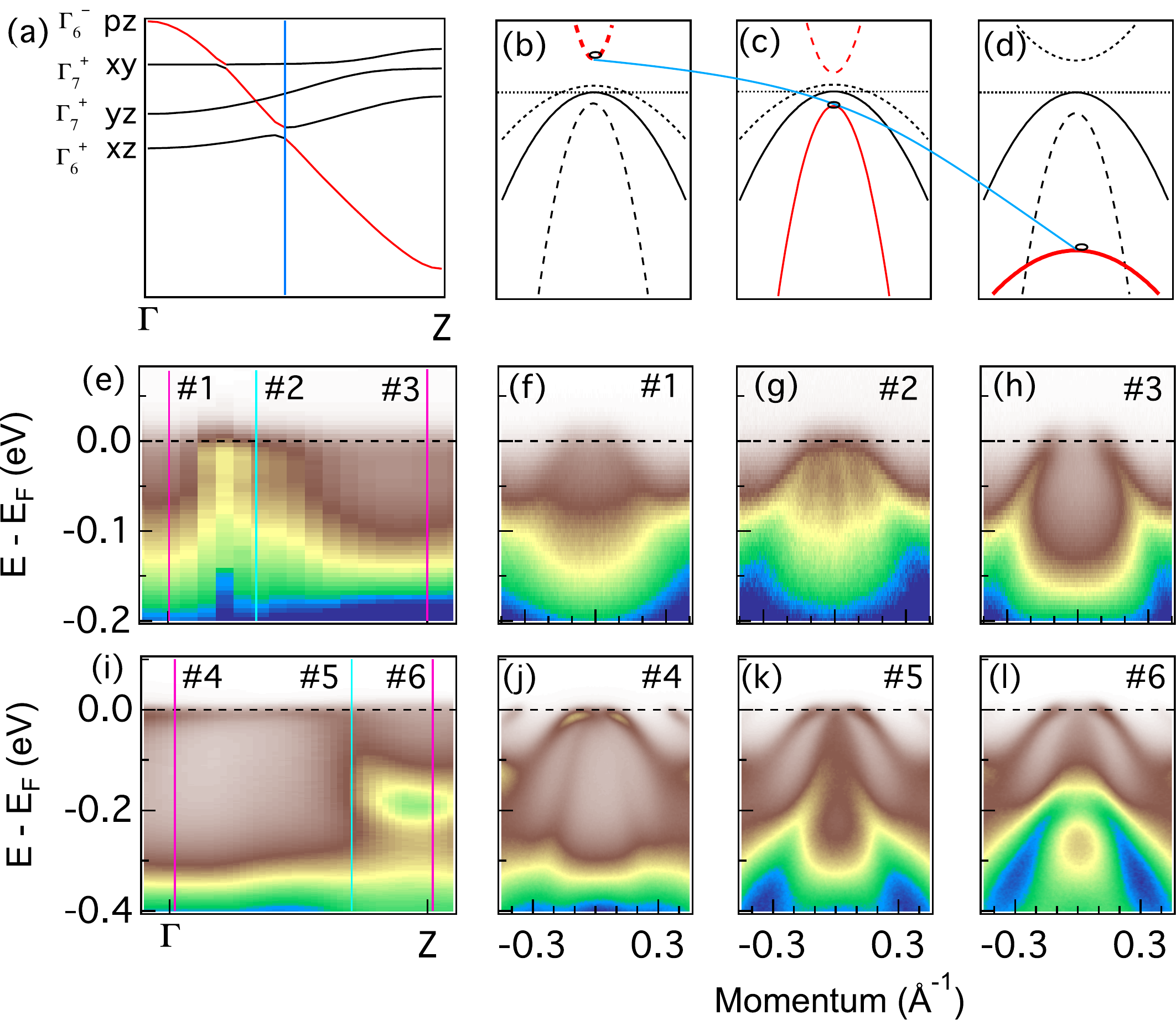}
 \caption{(Color online)
 Normal emission ARPES spectra. First row: schematic band dispersions. Second row: ARPES data on FeSe$_{0.5}$Te$_{0.5}$. Third row: ARPES data on LiFeAs. First column: dispersion at $k_x=0$ along the $\Gamma$-Z direction. Second, third and fourth columns: electronic dispersions along cuts indicated in the first column.
}\label{arp2}
\end{figure}

In the presence of SOC, the DFT calculations indicate that the $\Gamma_6^+$, $\Gamma_7^+$ and  $\Gamma_7^+$ states at the $\Gamma$ point mostly come from the $d_{xz}$, $d_{yz}$ and $d_{xy}$ orbitals, respectively, while the $\Gamma_6^-$ state (labeled as $\Gamma_2^-$ without SOC) have an important $p_z$ component besides the $d_{xy}$ orbital.
As shown in Fig.~\ref{arp2}(a), while the $p_z$ band locates above $E_F$ at $\Gamma$, it is shifted below $E_F$ upon moving along $\Gamma$-Z and it reaches its minimum at the Z point. On its way down, the $p_z$ band crosses the $d_{xy}$, $d_{yz}$ and $d_{xz}$ bands, opening a SOC gap with the $d_{xz}$ band. This situation is illustrated schematically in Fig.~\ref{arp2}(a). We now ask how the hybridization of the $p_z$ band with the $d_{xz}$ band should affect the ARPES measurements. Around the $\Gamma$ point, the $p_z$ band forms an electron-like band located above $E_F$ along the $\Gamma$-M direction. Therefore, it should not be observed in our experimental geometry. Due to selection rules, the $d_{xz}$ band should not be observed either and only the $d_{xy}$ and $d_{yz}$ band can possibly be seen, as shown schematically in Fig.~\ref{arp2}(b). However, because it hybridizes with the $p_z$ band, one should expect to be able to detect the $d_{xz}$ band near the crossing point below $E_F$, as illustrated schematically in Fig.~\ref{arp2}(c). Finally, our calculations indicate that the $p_z$ band has a hole-like dispersion near the Z point, and we should be able to observe it. On the other hand, the $d_{xz}$ band no longer hybridizes strongly with the $p_z$ band and its intensity should be significantly suppressed again in the $\sigma$ polarization, as described in Fig.~\ref{arp2}(d).

In order to confirm this dispersion, we compare in Fig.~\ref{arp2}(e) ARPES spectra recorded on FeSe$_{0.45}$Te$_{0.55}$ with different photon energies.
We see some intensity between the $\Gamma$ and Z points that we assign to the $p_z$ band sinking down from $\Gamma$ to Z. At the Z point, this band has merged with the strong $d_{z^2}$ band and it is thus undistinguishable. 
We display in Figs.~\ref{arp2}(f)-\ref{arp2}(h) three ARPES intensity cuts along $\Gamma$-M recorded on FeTe$_{0.55}$Se$_{0.45}$ and corresponding respectively to $k_z$ values around $\Gamma$ (Cut\#1, Fig.~\ref{arp2}(f)), the hybridization between the $p_z$ and $d_{xz}$ bands (Cut\#2, Fig.~\ref{arp2}(g)),
and around Z (Cut\#3, Fig.~\ref{arp2}(h)).
As the $d_{xy}$ band heavily renormalizes compared to the $d_{xz}/d_{yz}$ bands, it is  shallower and hardly resolved with weaker intensity~\cite{kfese2013}.
In Cut\#1 (Fig.~\ref{arp2}(f)), as expected, none of the $p_z$ and $d_{xz}$ bands are detected near $k_z = 0$. In contrast, not only the $d_{yz}$ band is observed in Cut\#2 (Fig.~\ref{arp2}(g)), but the $d_{xz}$ band as well due to its hybridization with the $p_z$ band. At the Z point, away from hybridization, the $d_{xz}$ band disappears from the ARPES spectrum (Fig.~\ref{arp2}(h)). Unfortunately, the $p_z$ band is too close to the strong $d_{z^2}$ band to be distinguished unambiguously.

As shown in Figs.~\ref{arp2}(i) to \ref{arp2}(l), the strong $k_z$ dispersion of the $p_z$ band is also observed in LiFeAs, which exhibits very clear spectral features. Because its bottom is located away from the $d_{z^2}$ band, the dispersion of the $p_z$ band along $\Gamma$-Z can be identified very clearly, as illustrated in Fig.~\ref{arp2}(i).
As with FeSe$_{0.45}$Te$_{0.55}$, the intensity of the $d_{xz}$ band is the strongest along Cut\#5 (Fig.~\ref{arp2}(k)), where it hybridizes with the $p_z$ band. Figs. \ref{arp2}(k)-\ref{arp2}(l) also show clearly that the $p_z$ band has a hole-like dispersion below $E_F$, as expected theoretically.

\begin{figure}[!tb]
 \includegraphics[clip,scale=0.65,angle=0]{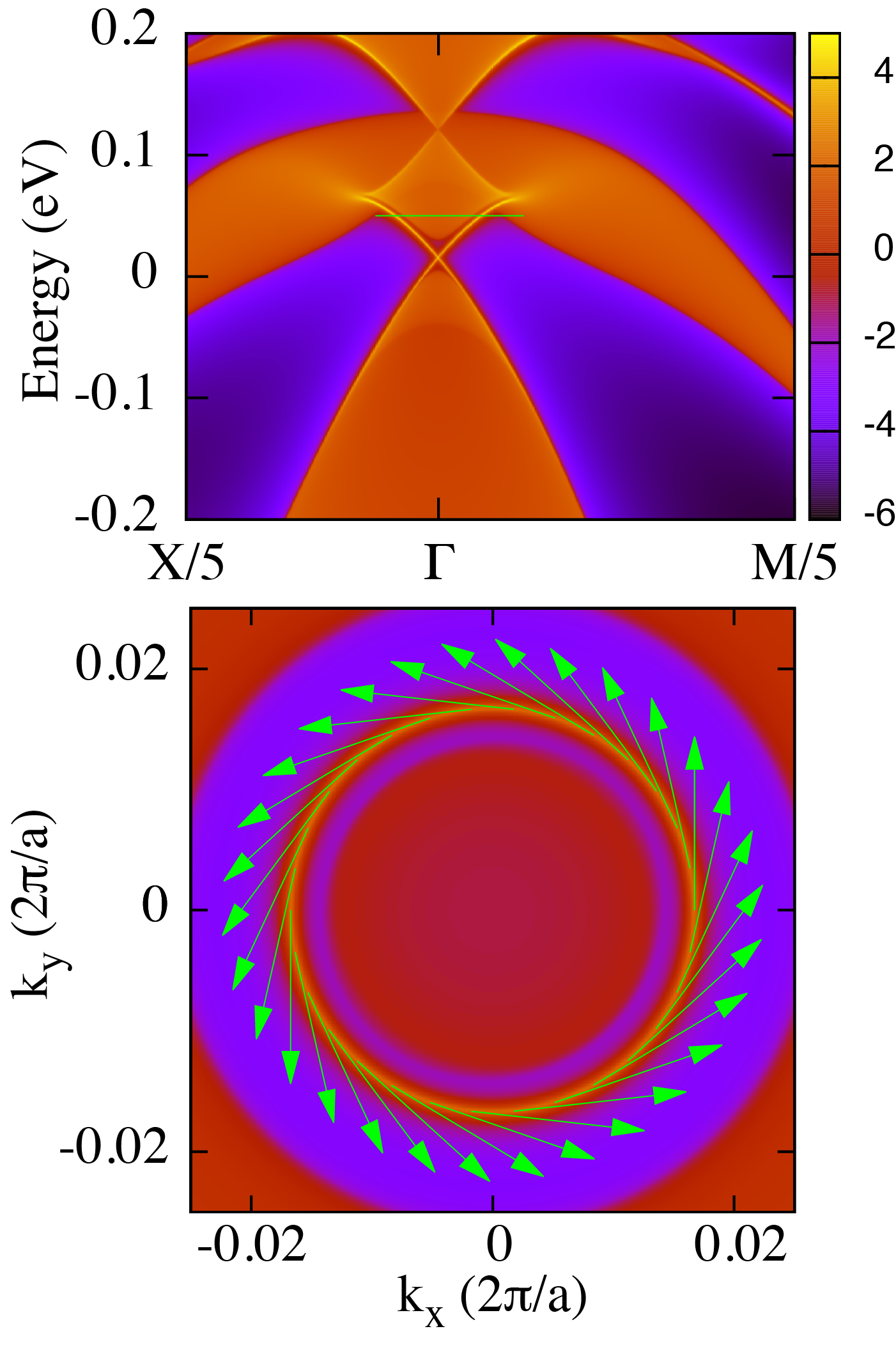}
  \caption{(Color online)
  Topological non-trivial 001 surface states and spin-resolved Fermi surface of these states.
  (a) Surface LDOS on the (001) surface for FeSe$_{0.5}$Te$_{0.5}$ considering DMFT on-site modification. (b) Fermi surface of the topological state (bright yellow circle beside the projected bulk states). The in-plane spin orientation is indicated by green arrows.
}
\label{surf}
\end{figure}


Finally, we analyze the spin-resolved Fermi surfaces around the Dirac point of the semi-infinite system formed by the SS to identify the non-trivial topology. With the surface Green's function calculated from the modified effective Hamiltonian considering DMFT on-site modification, the spin-filter surface states and the corresponding Fermi surfaces can be obtained directly.
From the dispersion of the SS shown in Fig.~\ref{surf}(a),
the protected SS emerges in the SOC gap around $E_F$ due to the non-trivial topological nature of the bulk system.
The Fermi surface of the SS ($E_F=50$ meV) is illustrated as a bright yellow circle in Fig.~\ref{surf}(b), and the spin orientation for the SS around the Fermi surface is marked by green arrows, which form a $\pi$ Berry phase enclosed.
The magnitude of the $z$ spin component is very small compared to the in-plane component. The $\pi$ Berry phase signifies the topological non-trivial properties of the bulk. 
By inducing a $s$-wave superconducting gap, the chiral SS can play an important role to produce the Majorana zero mode.


In conclusion, we have presented both theoretical and experimental evidences for a
topological non-trivial phase in FeSe$_{0.5}$Te$_{0.5}$. 
From the DFT calculations, 
we show that the band topology is sensitive to the intra-layer and inter-layer hopping terms, which can be tuned by the Te substitution. The Te substitution content $x$ strongly affects the structural, electronic and topological properties of these materials. 
We have identified the topologically non-trivial electronic band structure of FeSe$_{0.5}$Te$_{0.5}$ with a band inversion, characterized by an odd $\mathbb{Z}_2$ invariant and spin-moment locked SS.
Our ARPES data strongly support that the $\Gamma_2^-$ band forms a band inversion at the Z point.
The similar results can also be applied to iron-pnictides such as
LiFeAs. Due to the topological non-trivial surface states, the
FeSe$_{1-x}$Te$_x$ materials would be an ideal system for realizing
possible topological superconductor and Majorana fermions on the
surface.

We acknowledge discussions with J.P. Hu and experimental assistance
from Y.M. Xu. This work was supported by NSFC (11474340, 11274362,
11204359 and 11234014), the 973 program of China ( 2011CBA001000,
2011CBA00108 and 2013CB921700), and the ``Strategic Priority Research
Program (B)" of the Chinese Academy of Sciences (XDB07000000 and
XDB07020100).

\begin{widetext}
\begin{appendix}
\newpage
\section{Appendix A: Structural parameters of FeX and band structure of LiFeAs}
\label{app1}

\begin{table}[htb]
\centering
\caption{Structural parameters of PbO-structure FeX. The lattice parameters are
from experimental data~\cite{lisl}, and both optimized (Opt.) and experimental (Exp.) internal chalcogen positions z$_X$ are shown.
d$_z$ is the cartesian distance in the $z$ direction from the X plane to the Fe plane. }
\label{tab1}
\hspace{-1cm}
\begin{tabular}
[c]{cccccccccc}\hline\hline
 &a (\AA) &c (\AA)& z$_X$(Opt.)/d$_z$ (\AA)&z$_X$(Exp.)/d$_z$(\AA)\\
\hline
 FeSe & 3.784  & 5.502  & 0.2345/1.2902 & 0.2652/1.4591\\
 FeTe & 3.8215 & 6.2695 & 0.2496/1.5648 & 0.2829/1.7736\\
 FeSe$_{0.5}$Te$_{0.5}$& 3.7933 & 5.9552 & 0.2476/1.4745 & 0.2719/1.6192\\
\hline\hline
\label{structpara}
\end{tabular}
\end{table}

For LiFeAs, we used the experimental lattice parameters~\cite{liasstr} and relaxed the free internal coordinates. The experimental structure (space group P4/nmm, 129) has the Li sites in 2c positions, which lie above and below the centers of the Fe squares opposite the As. The calculated DFT band structure is given in Fig.~\ref{life}. The $\Gamma_2^-$ band shows a large dispersion and a band reversion along the $z$ direction.
\begin{figure}[htb]
\includegraphics[clip,scale=0.5,angle=0]{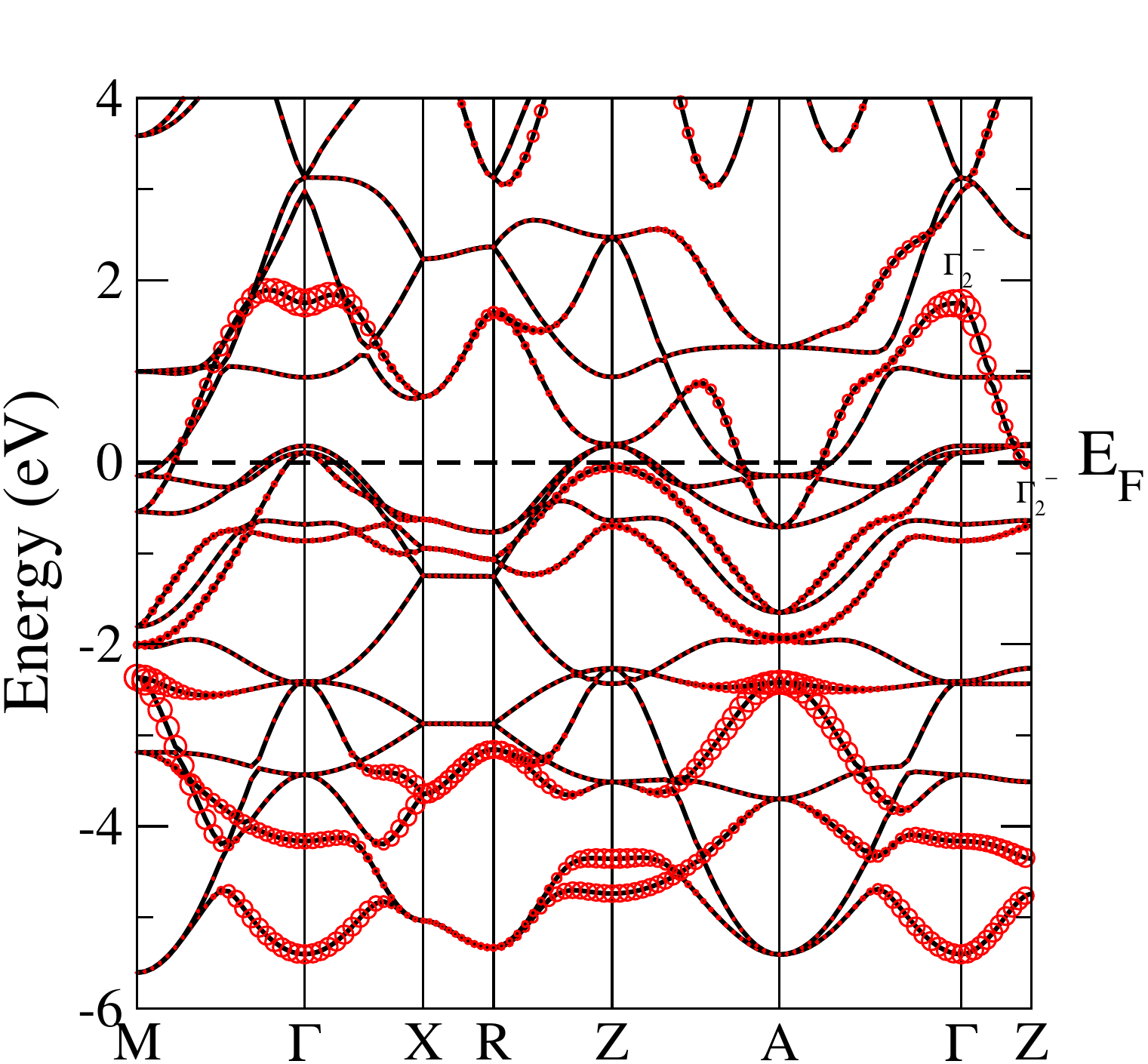}
  \caption{(Color online) Electronic band structure of LiFeAs. The $p_z$ component is highlighted by the size of the red circles. 
}\label{life}
\end{figure}

\section{Appendix B: Symmetry and parity analysis}
\label{symm}

Since the system has inversion symmetry, it is convenient to combine these orbitals into bonding and antibonding states with definite parity as
\begin{eqnarray*}
 |D_\alpha^{\pm}\rangle&=&\frac{1}{\sqrt{2}}(|Fe_{\alpha}\rangle\pm|Fe'_{\alpha}\rangle),\\
 |P_\beta^{\pm}\rangle&=&\frac{1}{\sqrt{2}}(|X_{\beta} \rangle\mp|X'_{\beta}\rangle)
\end{eqnarray*}
Using the orbitals with definite parity defined in the main text, we can label all the non-spin-orbital (NSO) bands with the irreducible representations (IRs), which are given in Table~\ref{chartsymb}.
The $\Gamma_2^-$ band is composed of the antibonding states with both $d_{xy}$ and $p_z$ characters.
 At both the $\Gamma$ and Z points, the IRs of the D$_{4h}$ group are labeled as $\Gamma_n^\pm$, whereas along the $\Gamma$-Z high-symmetry line, we label these IRs as $\Lambda_n$ because of its C$_{4v}$ symmetry.
In our conventions, the ($x$,$y$) axes are rotated by 45$^\circ$
as compared to the crystallographic axes,
so that the $d_{xy}$ orbital in our definition is the one
pointing from Fe to chalcogen atoms, as shown in Fig.~1.

\begin{table}[htb]
\centering
\caption{Combined orbitals and related irreducible representations}
\hspace{-1cm}
\begin{tabular}
[c]{cccccc|cc}\hline\hline
D/P &(+/-) & z$^2$ & yz/xz & xy & x$^2$-y$^2$ &x/y& z\\
\hline
D$_{4h}$&+ & $\Gamma_1^+$ & $\Gamma_5^+$ & $\Gamma_4^+$ & $\Gamma_3^+$ & $\Gamma_5^+$ & $\Gamma_1^+$\\
        &- & $\Gamma_3^-$ & $\Gamma_5^-$ & {\color{red}$\Gamma_2^-$} & $\Gamma_1^-$ & $\Gamma_5^-$ & {\color{red}$\Gamma_2^-$} \\
\hline
C$_{4v}$&+ & $\Lambda_1$  & $\Lambda_5$  & $\Lambda_4$  & $\Lambda_3$ & $\Lambda_5$ & $\Lambda_1$\\
        &- & $\Lambda_4$  & $\Lambda_5$  & {\color{red}$\Lambda_1$}  & $\Lambda_2$ & $\Lambda_5$ & {\color{red}$\Lambda_1$} \\
\hline\hline
\label{chartsymb}
\end{tabular}
\end{table}

\begin{figure}[htb]
\includegraphics[clip,scale=0.5,angle=0]{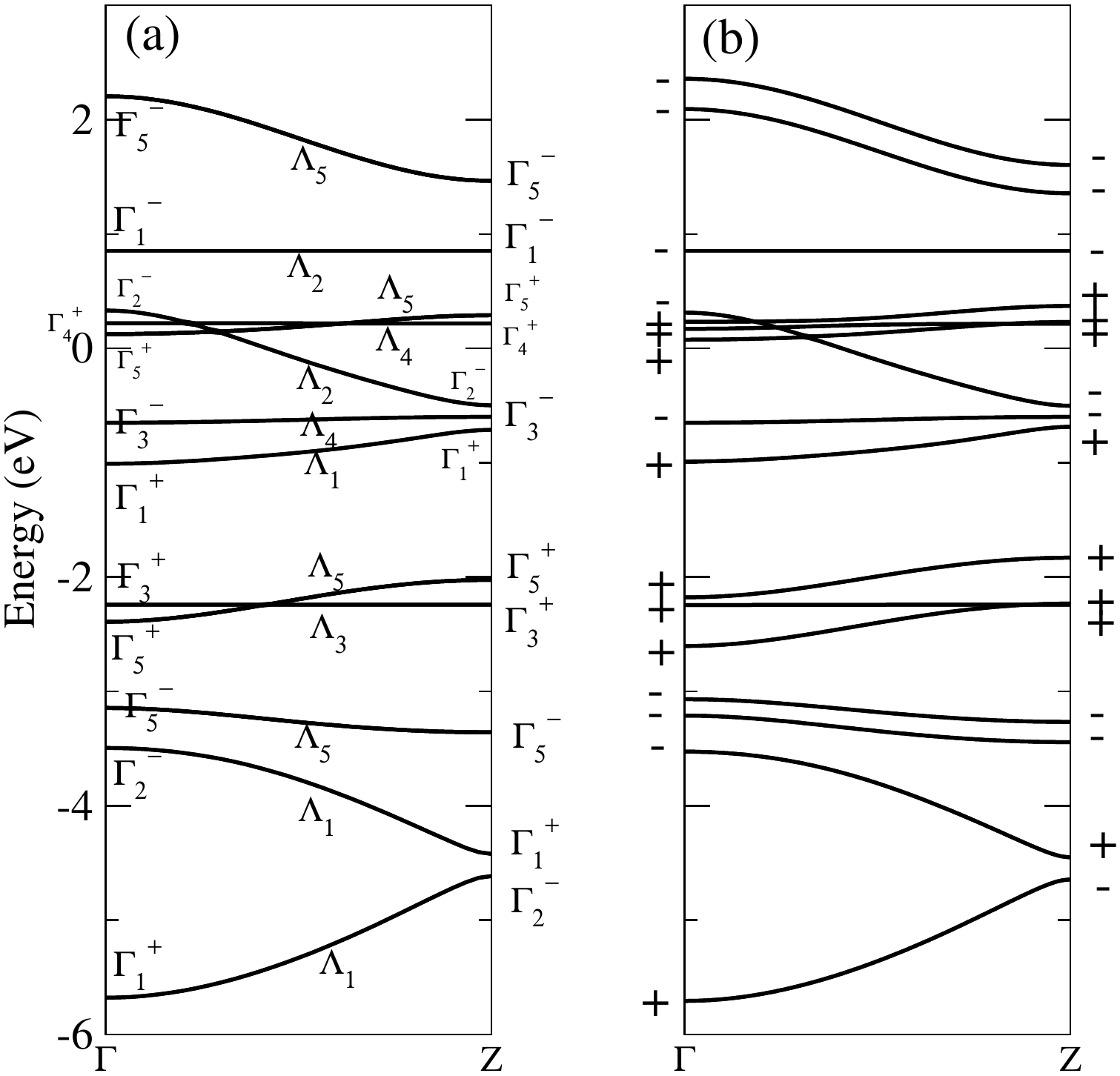}
  \caption{(Color online) Electronic band structure along $\Gamma$-Z.
  (a) The IRs are labeled in the NSO bands.
  (b) The parities are labeled in the SOC bands.
}\label{gz}
\end{figure}

The IR labels along the $\Gamma$-Z line are shown in Fig.~\ref{gz}(a). Around $E_F$, the $D_{xz/yz}^+$ bands can mix with the $P_{x/y}^+$ orbitals
with the same $\Gamma_5^+$ representations, although the P-D hybridization is not strong due to the little orbital overlap in the real space.
For the $\Gamma_2^-$ representation, the $D_{xy}^-$ character can mix with the $P_{z}^-$ band,
and the $d_{xy}$ and $p_z$ orbitals have a strong hybridization along the $z$ direction. However, $d_z$ becomes larger with the Te substitution content $x$ increasing, thus leading to weaker hybridization. As a result, the $\Gamma_2^-$ band, consisting of $d_{xy}$ and $p_z$ orbitals,
sinks down to E$_F$ and has a large dispersion in FeSe$_{0.5}$Te$_{0.5}$. Moreover, due to selection rules, the bonding state of $D_{xy}^+$ cannot mix with the $p_z$ band, including along the $\Gamma$-Z line. As a consequence, the  $\Gamma_4^+$ band characterized by the $D_{xy}^+$ representation exhibits a weak dispersion in the $z$ direction, which is not sensitive to the height d$_z$.
Beside, all the IRs in NSO bands and parities in SOC bands are presented in Table~\ref{bandsirrep}.
``$(+,-)$'' denotes that the quadruple bands consist of two Kramers pairs with opposite parity.
The non-trivial $\mathbb Z_2$ index is presented beside the vertical line, 
which denotes the hypothetical Fermi level in the main text.

\begin{center}
\begin{table}[h]
\centering
\caption{The D$_{4h}$ IRs and parities for each band. }
\label{tab2}
\hspace{-1cm}
\begin{tabular}
[c]{ccccccccccc|cccccc}
\hline\hline
IRs(NSO)&&&P&&&&&&&D&&&&&&\\
$\Gamma$&$\Gamma_1^+$&$\Gamma_2^-$&$\Gamma_5^-$&&$\Gamma_5^+$&&
    $\Gamma_3^+$&$\Gamma_1^+$&$\Gamma_3^-$&$\Gamma_5^+$&&$\Gamma_4^+$&
    $\Gamma_2^-$&$\Gamma_1^-$&$\Gamma_5^-$&\\
Z&$\Gamma_2^-$&$\Gamma_1^+$&$\Gamma_5^-$&&$\Gamma_3^+$&
    $\Gamma_5^+$&&$\Gamma_1^+$&$\Gamma_3^-$&$\Gamma_2^-$&$\Gamma_4^+$&
    $\Gamma_5^+$&&$\Gamma_1^-$&$\Gamma_5^-$&\\
 \hline
Parities(SO)&1&2&3&4&5&6&7&8&9&10&11&12&13&14&15&16\\
\hline
$\Gamma$&+&-&-&-&+&+&+&+&-&+&+&+&-&-&-&-\\
Z&-&+&-&-&+&+&+&+&-&-&+&+&+&-&-&-\\
X&(+&-)&(+&-)&(+&-)&(+&-)&(+&-)&(+&-)&(+&-)&(+&-)\\
M&(+&-)&(+&-)&(+&-)&(+&-)&(+&-)&(+&-)&(+&-)&(+&-)\\
R&(+&-)&(+&-)&(+&-)&(+&-)&(+&-)&(+&-)&(+&-)&(+&-)\\
A&(+&-)&(+&-)&(+&-)&(+&-)&(+&-)&(+&-)&(+&-)&(+&-)\\
\hline
Z$_2$&&&&&&&&&&1&&&&&&\\
\hline\hline
\label{bandsirrep}
\end{tabular}
\end{table}
\end{center}

\section{Appendix C: LDA+DMFT confirmation }
\label{dmftcal}
The LDA + DMFT
method has proven to be a powerful technique to study the
electronic structure of correlated systems. In this
section, we apply this method to  FeSe$_{0.5}$Te$_{0.5}$
with a local Coulomb integral U=F$_0$=4.0 eV and a Hund's coupling J=0.7 eV, and confirm that it has a topological character of band inversion by computing the
topological invariants within the DMFT framework.

The one-electon spectral function is defined as
\begin{eqnarray*}
A_{\mathbf k}(\omega)=-\frac{1}{\pi}\frac{Im\Sigma(\omega)}
{(\omega +\mu -\epsilon_{\mathbf k}-Re\Sigma(\omega))^2+Im\Sigma(\omega)^2}
\end{eqnarray*}
in terms of the LDA band dispersion $\epsilon_{\mathbf k}$
and the self-energy $\Sigma(\omega)$.
This momentum-resolved spectra $A_{\mathbf k}(\omega)$ is shown in the upper panel of
Fig.~\ref{akw}, where the overall renormalization of the bands
and the bandwidth reduction are apparent.
However, the relative positions of different bands almost do not change. The $\Gamma_2^-$ and $\Gamma_5^+$ bands crossing near $E_F$ remains in the correlated system.

\begin{figure}[h]
\includegraphics[clip,scale=0.6,angle=270]{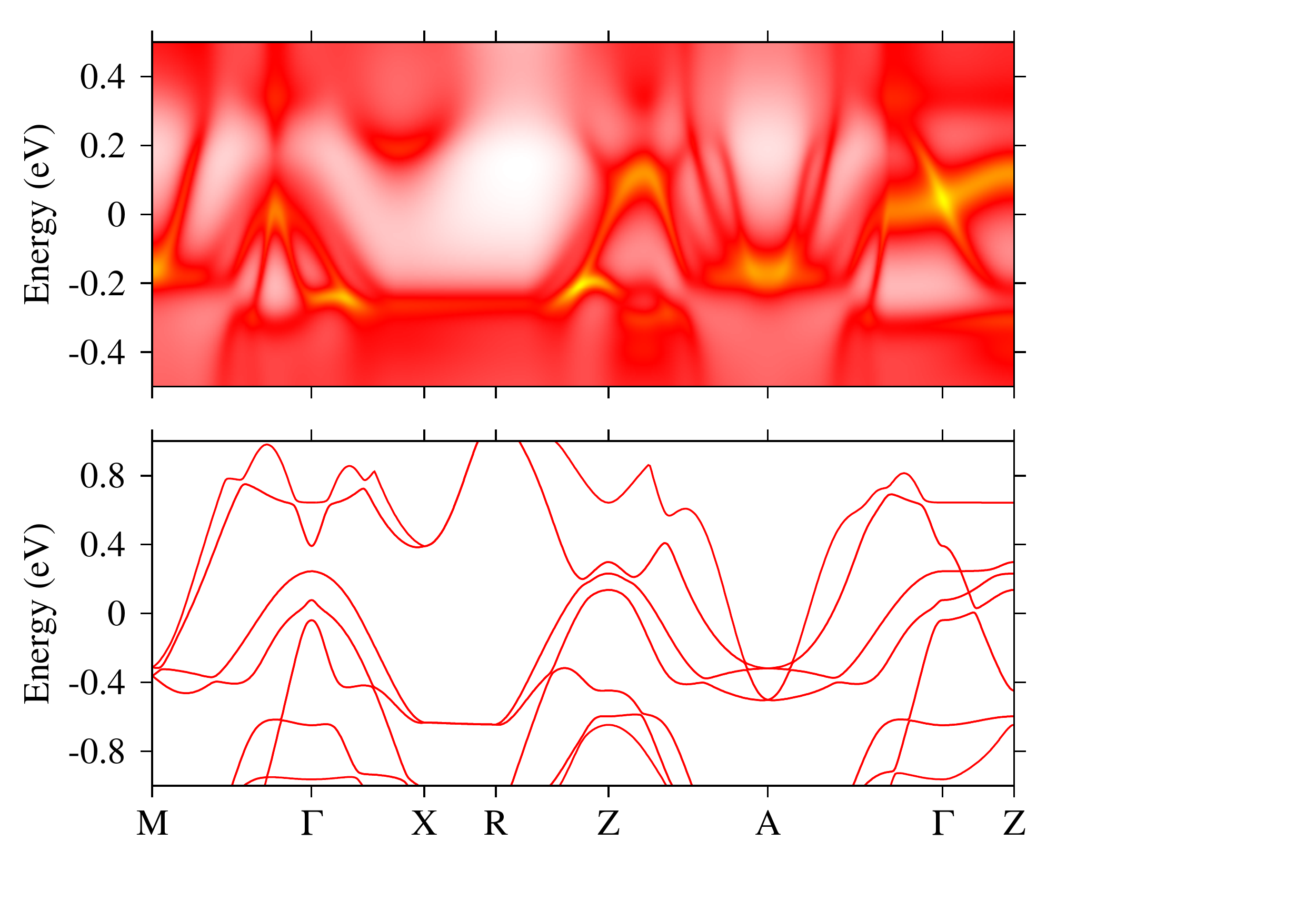}
  \caption{(Color online)
  (Upper panel) LDA+DMFT calculations.
  (Lower panel) SOC band structure only taking the LDA+DMFT on-site
  modification into consideration.}
\label{akw}
\end{figure}

In a previous work~\cite{zwang1}, only the Green's function at zero frequency $G_{\mathbf k}^{-1}(0)=\mu-\epsilon_{\mathbf k} -\Sigma_{\mathbf k}(0)$
is needed to determine the topology of the quasiparticle states, since renormalization does not change the non-trivial topological nature of the bulk system.
Following Refs.~\cite{zwang2}, we compute the topological invariant of an interacting system with the Hamiltonian defined by $H_t({\mathbf k, \omega})= H({\mathbf k}, \omega)+\Sigma_{\mathbf k}(0)-\mu$.
A modified effective dispersion of
$H_t({\mathbf k})$ for which we neglected the imaginary part of $\Sigma_{\mathbf k}(0)$ is given in the lower panel of Fig.~\ref{akw}.
By introducing correlations, the bandwidth is reduced by approximately half, but the band inversion character still exists. According to the parity criteria, we conclude that FeSe$_{0.5}$Te$_{0.5}$ remains a topological correlated system. The topological surface state is derived from this modified effective TB Hamiltonian.

\end{appendix}

\end{widetext}

\end{document}